%% file: jane.tex
\documentclass{svproc}
\usepackage{url}

\input{macros}

\begin{document}
\mainmatter              
%


\title{Joint Use of Node Attributes and Proximity \\ for Node Classification}
\titlerunning{Joint Use of Node Attributes and Proximity for Node Classification}

\author{Arpit Merchant      \and
        Michael Mathioudakis 
}
\tocauthor{Arpit Merchant, Michael Mathioudakis}

\institute{University of Helsinki, Helsinki, Finland \\
          \email{firstname.lastname@helsinki.fi} \\
}

%
%
%

\maketitle              


\input{abstract}

\keywords{node classification, graph embeddings, spectral graph analysis}

\input{introduction}

\input{setting}

\input{our_approach}

\input{experiments_synthetic_data}

\input{experiments_real_data}

\input{conclusion}


\bibliographystyle{splncs03}      
\bibliography{references}

\end{document}

%% file: macros.tex
\usepackage{graphicx}
\usepackage{textcomp}
\usepackage{xcolor}

\usepackage{verbatim,paralist,subscript,bbold,bbm,microtype}

\usepackage{url}
\usepackage[lined, boxed, ruled, commentsnumbered]{algorithm2e}
\usepackage{mathtools,array,tikz,savesym}
\usepackage{amsfonts}
\usepackage{nicefrac,xfrac}
\usepackage{booktabs,dcolumn,tabularx}

\usepackage{etoolbox}

\usepackage{multirow,caption,subcaption}
\usepackage{xspace}
\usepackage{cancel}
\usepackage{hyperref}

\usetikzlibrary{arrows,shapes}

\def\BibTeX{{\rm B\kern-.05em{\sc i\kern-.025em b}\kern-.08em
    T\kern-.1667em\lower.7ex\hbox{E}\kern-.125emX}}

\savesymbol{u}
\savesymbol{y}
\savesymbol{U}
\savesymbol{a}
\savesymbol{t}
\savesymbol{L}
\savesymbol{do}

\DeclareMathOperator*{\argmin}{arg\,min}
\DeclareMathOperator*{\argmax}{arg\,max}

\renewcommand{\algorithmcfname}{Algorithm}
\newcommand{\R}{\mathbb{R}}

\newcommand{\norm}[1]{\left\lVert#1\right\rVert}

\newcommand{\round}[1]{\left(#1\right)}
\newcommand{\parenthesis}[1]{(#1)}
\newcommand{\braces}[1]{\left\{#1\right\}}
\newcommand{\squares}[1]{\left[#1\right]}
\newcommand{\indexset}[1]{\{1,\ldots,{#1}\}}

\newcommand{\algo}{\textsc{JANE}\xspace}

\newcommand{\prob}[1]{\text{P}\squares{#1}}

\newcommand{\spara}[1]{\smallskip\noindent{\bf{#1}}}

\newcommand{\squishlist}{\begin{list}{$\bullet$}
  {\setlength{\itemsep}{0pt}
    \setlength{\parsep}{3pt}
    \setlength{\topsep}{3pt}
    \setlength{\partopsep}{0pt}
    \setlength{\leftmargin}{1.5em}
    \setlength{\labelwidth}{1em}
    \setlength{\labelsep}{0.5em}}}
\newcommand{\squishend}{
\end{list}}

\newcommand{\hide}[1]{}

\newcommand{\graph}{\ensuremath{\mathcal{G}}}
\newcommand{\vertexset}{\ensuremath{\mathcal{V}}}
\newcommand{\edgeset}{\ensuremath{\mathcal{E}}}

\newcommand{\scale}{\ensuremath{s}\xspace}
\newcommand{\softmax}{\ensuremath{\sigma}\xspace}
\newcommand{\likelihood}{\ensuremath{\mathcal{L}}}

\newcommand{\A}{\ensuremath{{\mathbf{A}}}\xspace}
\newcommand{\diagmat}{\ensuremath{{\mathbf{D}}}\xspace}
\newcommand{\laplacian}{\ensuremath{{\mathbf{L}}}\xspace}
\newcommand{\adj}[2]{\ensuremath{a_{#1#2}}\xspace}
\newcommand{\eigval}[2]{\ensuremath{\lambda_{#1}\round{#2}}\xspace}

\newcommand{\eigvecExp}[2]{\ensuremath{{\mathbf{e}}_{#1}\round{#2}}\xspace}

\newcommand{\X}{\ensuremath{{\mathbf{X}}}\xspace}
\newcommand{\x}[1]{\ensuremath{{\mathbf{x}}_{#1}}\xspace}
\newcommand{\U}{\ensuremath{{\mathbf{U}}}\xspace}
\newcommand{\u}[1]{\ensuremath{{\mathbf{u}}_{#1}}\xspace}
\newcommand{\labelset}{\ensuremath{{\mathbf{M}}}\xspace}
\newcommand{\Y}{\ensuremath{{\mathbf{Y}}}\xspace}
\newcommand{\y}[1]{\ensuremath{{\mathbf{y}}_{#1}}\xspace}

\newcommand{\paramshat}{\ensuremath{\hat{{\theta}}}\xspace}
\newcommand{\Uhat}{\ensuremath{\hat{{\mathbf{U}}}}\xspace}
\newcommand{\uhat}[1]{\ensuremath{\hat{{\mathbf{u}}_{#1}}}\xspace}

\newcommand{\inedgeset}{\round{i,j} \in \edgeset}
\newcommand{\notinedgeset}{\round{i,j} \notin \edgeset}
\newcommand{\normuiuj}{\norm{\u{i} - \u{j}}^2}

\newcommand{\unb}[1]{\ensuremath{{u}_{#1}}}
\newcommand{\normuiujnb}{\norm{\unb{i} - \unb{j}}^2}
\newcommand{\normuiujnbsigma}{\frac{\norm{\unb{i} - \unb{j}}^2}{\scale^2}}

\newcommand{\expuiujnbsigma}{{e}^{\frac{-\normuiujnb}{\scale^2}}}

\newcommand{\W}[1]{\ensuremath{{W}^{\round{#1}}}\xspace}
\newcommand{\Wh}[1]{\ensuremath{\hat{W}^{\round{#1}}}\xspace}
\newcommand{\WW}{\ensuremath{\mathbf{W}}\xspace}
\newcommand{\What}{\ensuremath{\hat{\mathbf{W}}}\xspace}

\newcommand{\data}{\ensuremath{\mathcal{D}}\xspace}
\newcommand{\params}{\ensuremath{\mathcal{\theta}}\xspace}
\newcommand{\loss}{\ensuremath{\mathcal{C}}}
\newcommand{\a}[2]{{a}_{#1}^{\round{#2}}}
\newcommand{\z}[2]{{z}_{#1}^{\round{#2}}}
\newcommand{\doh}[2]{\frac{\partial #1}{\partial #2}}

%% file: abstract.tex
\begin{abstract}

The task of node classification is to infer unknown node labels, given the labels for some of the nodes along with the network structure and other node attributes.
Typically, approaches for this task assume homophily, whereby neighboring nodes have similar attributes and a node's label can be predicted from the labels of its neighbors or other proximate (i.e., nearby) nodes in the network.
However, such an assumption may not always hold -- in fact, there are cases where labels are better predicted from the individual attributes of each node rather than the labels of its proximate nodes.
Ideally, node classification methods should flexibly adapt to a range of settings wherein unknown labels are predicted either from labels of proximate nodes, or individual node attributes, or partly both.
In this paper, we propose a principled approach, JANE, based on a generative probabilistic model that \underline{j}ointly weighs the role of \underline{a}ttributes and \underline{n}ode proximity via \underline{e}mbeddings in predicting labels. 
Our experiments on a variety of network datasets demonstrate that JANE exhibits the desired combination of versatility and competitive performance compared to standard baselines. 

\end{abstract}

		

%% file: introduction.tex
\section{Introduction} \label{sec:introduction}

Semi-supervised node classification over attributed graphs involves a graph with known structure, where each node is associated with a label (or `class', a categorical variable), as well as other attributes.
However, labels are known only for some of the nodes and are unknown for others.
Given the graph structure, the attributes of all nodes, and the labels of some of the nodes, the goal is to predict the labels for the remaining nodes.
This task finds application in many domains such as information networks \cite{sen2008collective}, complex systems \cite{zhu2009introduction}, and protein function identification \cite{Hamilton2017a}.

\begin{figure*}[t!]
    \captionsetup[subfigure]{font=scriptsize,labelfont=scriptsize}
    \centering
    \begin{subfigure}[t]{0.32\textwidth}
        \includegraphics[width=\textwidth]{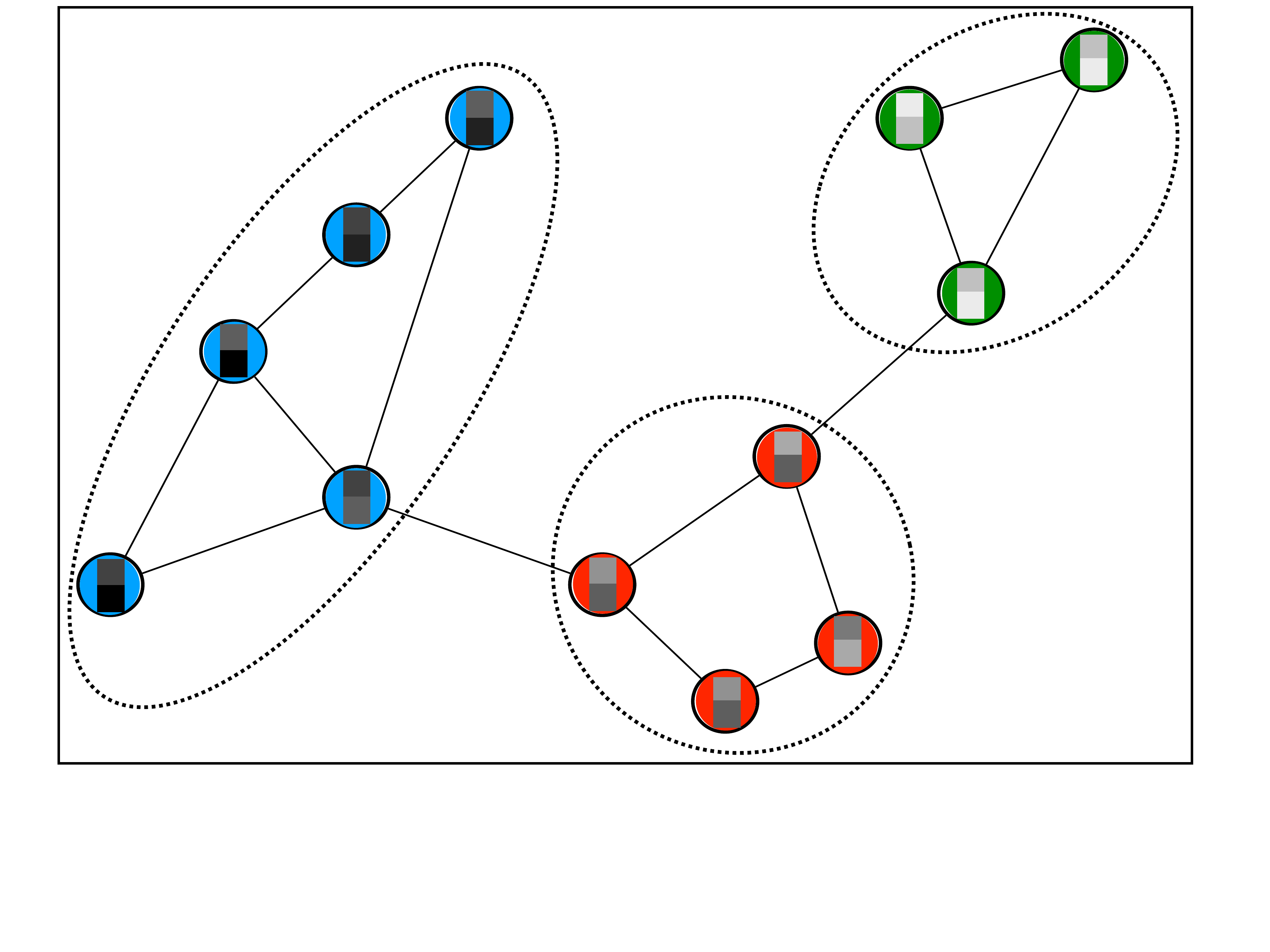}
        \caption{Nodes with similar labels are adjacent to each other and have similar features.}
        \label{subfig:XA_match}
    \end{subfigure}
    \begin{subfigure}[t]{0.32\textwidth}
        \includegraphics[width=\textwidth]{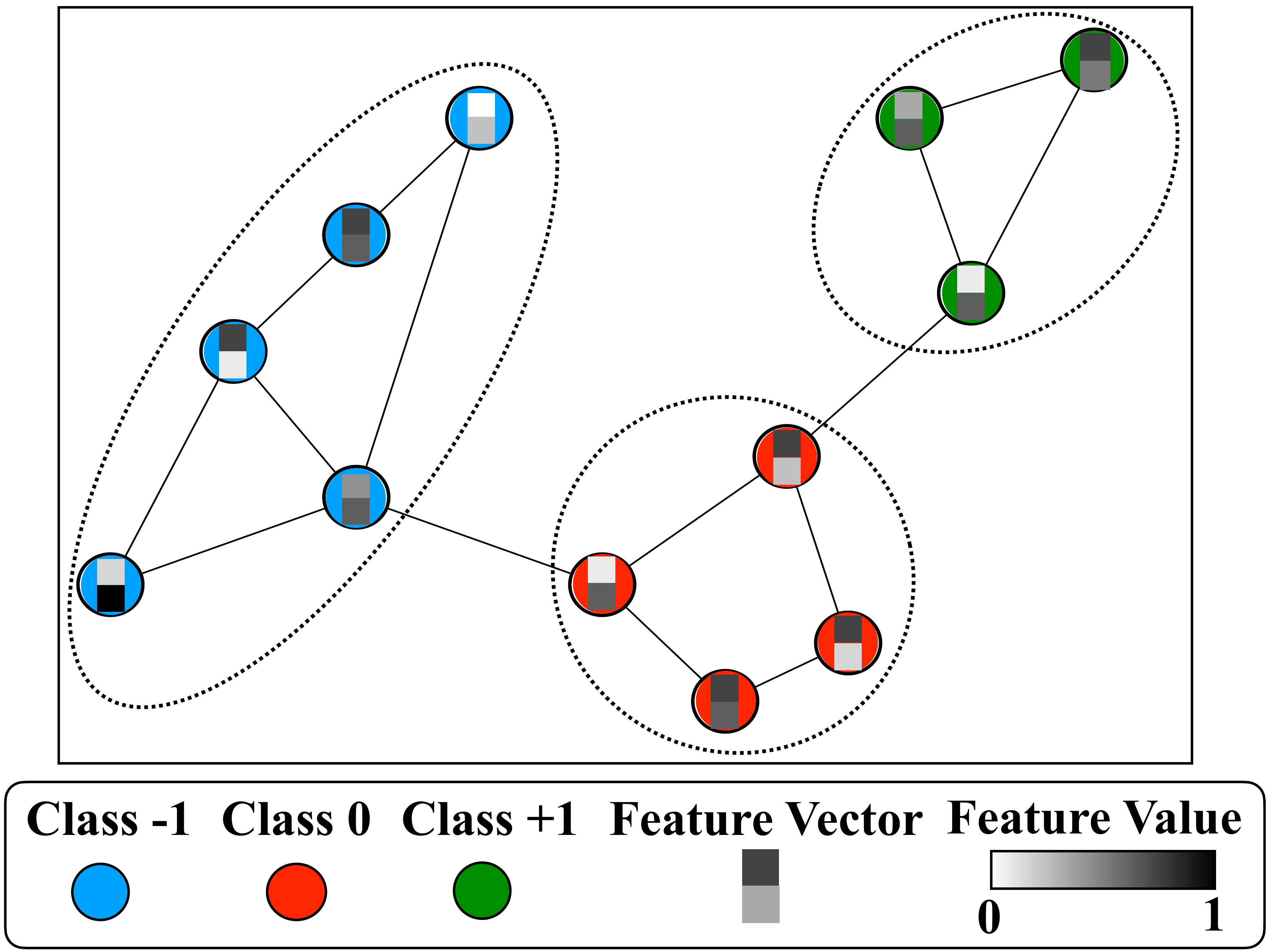}
        \caption{Nodes with similar labels are adjacent to each other but have different features.}
        \label{subfig:A_match}
    \end{subfigure}
    \begin{subfigure}[t]{0.32\textwidth}
        \includegraphics[width=\textwidth]{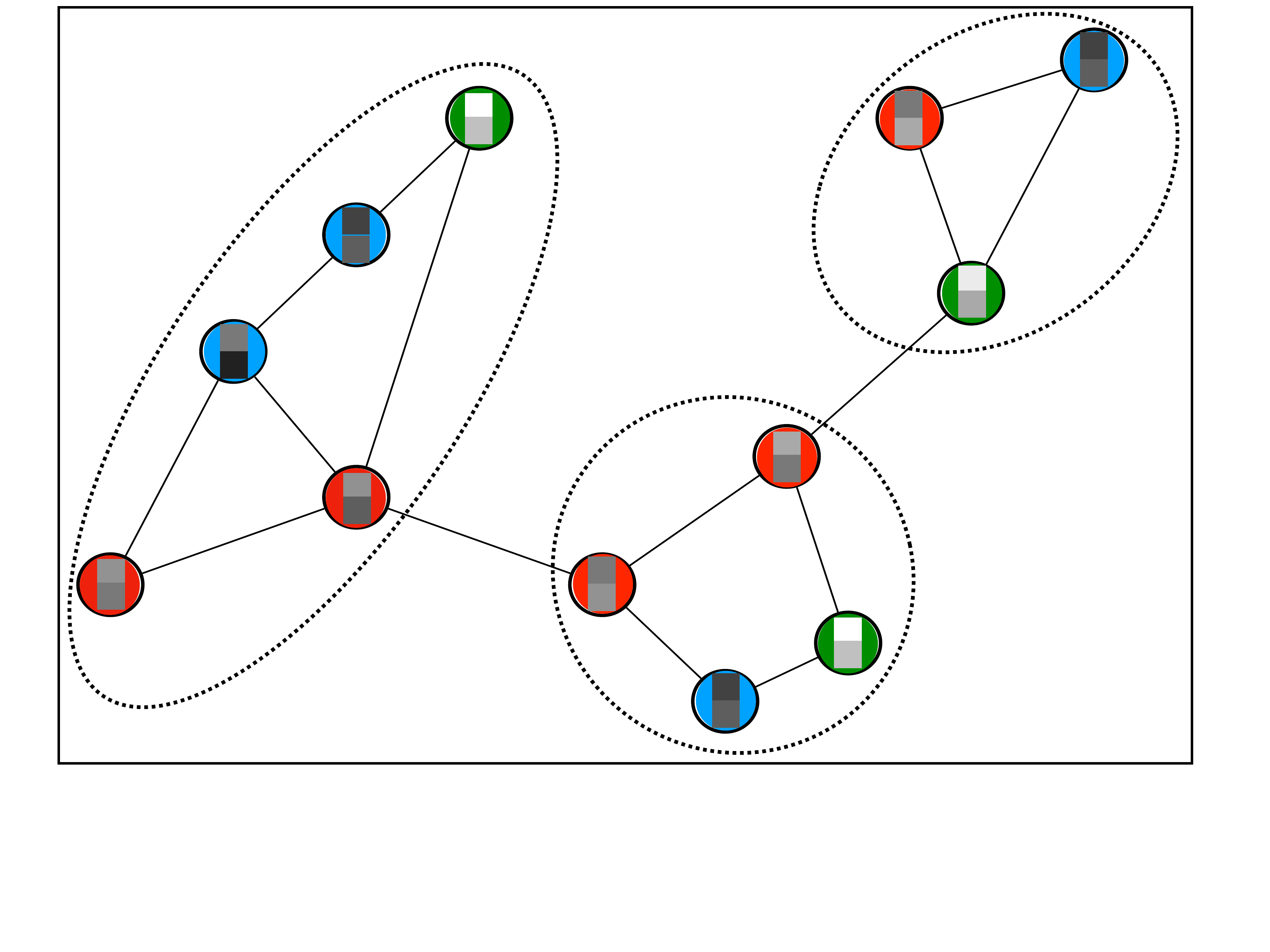}
        \caption{Nodes with similar labels have similar features but are not adjacent to each other.}
        \label{subfig:X_match}
    \end{subfigure}
    \caption{Node labels may correlate with either graph structure (\ref{subfig:A_match}), or node attributes(\ref{subfig:X_match}), or both (\ref{subfig:XA_match}). GCNs would perform well in the case of Fig.~\ref{subfig:XA_match}, ~\ref{subfig:A_match}, but not in Fig.~\ref{subfig:X_match} because they predict the label of a node by aggregating features of adjacent nodes. Conversely, attribute-based models would perform well in the case of Fig~\ref{subfig:XA_match}, ~\ref{subfig:X_match}, but not in Fig.~\ref{subfig:A_match} because they do not exploit the correlation between adjacent nodes and their labels.}
    \label{fig:toy_graph_illustration}
  \end{figure*}

\spara{Previous Work.} Motivated by theories of homophily \cite{mcpherson2001birds} and social influence \cite{marsden1993network}, a common assumption is that adjacent nodes tend to have similar labels. 
For instance, in a social network, friends are likely to vote for the same political party. 
Approaches that rely on this assumption typically enforce homophily by considering node proximity and assigning the same label to nearby nodes.
In label propagation, for example, labels diffuse from labeled nodes to their unlabeled neighbors in an iterative manner until convergence~\cite{zhu2002learning}. Other methods induce label uniformity within cuts or clusters of the graph~\cite{blum2004semi,joachims2003transductive,belkin2006manifold}; or consider node proximity in a latent space that preserves graph distances, as in DeepWalk~\cite{Perozzi2014} and similar matrix factorization approaches~\cite{Qiu2017}. 

However, the aforementioned methods ignore other node attributes, which can be detrimental.
For example, Hamilton et al.~\cite{Hamilton2017a} show that, for certain predictive tasks on citation and social graphs, a linear classifier that is built only on node attributes outperforms approaches such as DeepWalk that are based on node proximity but ignore node attributes. 
Moreover, while homophily is often observed in some classification tasks, it is not uncommon to find that adjacent nodes do not share a particular label and that, in such cases, other node attributes can serve as better label predictors than graph structure~\cite{lorrain1971structural,sailer1978structural}. 
For example, two individuals may be friends (i.e., be connected on a social network) but vote for different political parties (`label') -- something that could be better predicted by rich, node-level attribute data (e.g., geographic location, income, or profession).
Therefore, it is important to appropriately leverage both the graph structure and node attributes, for label prediction. 

Partially addressing this limitation, AANE~\cite{huang2017accelerated} and DANE~\cite{gao2018deep} combine low-dimensional encodings of node attributes with graph-distance-preserving node embeddings, and use them as input features for label prediction.
However, they do not account for known labels during training, thus potentially ignoring information that would be useful in predicting the unknown labels.
LANE~\cite{huang2017label} overcomes this limitation, by learning a the joint latent representations of node attributes, proximity, and labels.  
However, LANE does not directly address the node classification task, i.e., it does not optimize the conditional probability distribution of node labels given the node attributes and graph structure, but rather targets their joint distribution of all quantities. 

Separately, graph convolutional networks use graph topology for low-pass filtering on node features \cite{wu2019net}. GAT \cite{velivckovic2017graph} introduce an attention mechanism to learn weights and aggregate features. GraphSAGE \cite{Hamilton2017a} use mean/max pooling to sample and aggregate features from nodes' local neighbourhoods. However, these convolutions are equivalent to repeated smoothing over the node attributes and performance quickly degrades \cite{Li2018}. Subsequent approaches such as DiffPool \cite{Ying2018} have sought to address this limitation, but these too aggressively enforce homophily and require that nodes with the same labels have similar graph and attribute representations. Recent work by AM-GCN \cite{wang2020gcn} attempts to weaken this assumption by analyzing the fusing capabilities of convolutional models. They define two modules, one each for the topology space and feature space, and adaptively combine them using an attention mechanism.

\spara{Our contribution.} 
In this work, we develop an approach to node classification that is able to flexibly adapt to and perform consistently well in a range of settings, from cases where node labels exhibit strong homophily (i.e., a node's label can be determined by the labels of its neighbors) to cases where labels are independent of graph structure and solely determined by other node attributes, as well as cases that lie between these two extremes.
We propose a novel and principled approach to node classification based on a generative probabilistic model that jointly captures the role of graph structure and node similarity in predicting labels. 
Our analysis leads to \algo, an algorithm that identifies a maximum-likelihood estimate of the model that can be employed to perform node classification.
Unlike aforementioned approaches that are heavily based on label homophily (e.g., label propagation), we account not only for the graph structure but also for node attributes, and flexibly weigh each of them appropriately depending on the case.
Moreover, unlike AANE~\cite{huang2017accelerated} and DANE~\cite{gao2018deep}, our approach learns a low-dimensional node representation informed by labels, that is then used for prediction.
And unlike LANE~\cite{huang2017label}, we directly optimize the conditional probability of labels given the graph structure and node attributes without necessarily enforcing homophily.

We summarize our main contributions below:
\squishlist
    \item Define a formal generative framework tailored to node classification, that captures the relationship between graph structure, node attributes, and node labels. 
    \item Develop \algo, an algorithm for semi-supervised node classification, based on maximum likelihood estimate (MLE) of the model parameters.
    \item Demonstrate shortcomings of existing approaches and versatility of \algo on synthetic datasets.
    \item Empirically validate \algo's performance on benchmark, real datasets compared to standard baselines. 
\squishend

%% file: setting.tex
\section{Problem Setting} \label{sec:setting}

Let us consider an undirected and connected graph $\graph = \round{\vertexset, \edgeset}$ of node size $|\vertexset| = n$. Let its structure be represented by the adjacency matrix $\A = \squares{\adj{i}{j}} \in \R^{n \times n}$.
Denote $\diagmat = \text{diag} \round{d_1, d_2,\ldots,d_n}$ to be the degree matrix where $d_i = \sum_{j} \adj{i}{j}$; and $\laplacian = \diagmat - \A$ as its unnormalized Laplacian matrix. 
Let $\eigval{i}{\laplacian}$ be the $i$-th smallest eigenvalue of $\laplacian$ and $\eigvecExp{i}{\laplacian}$ its corresponding eigenvector. 

Each node in the graph is associated with the following: $d$ observed attributes $\x{} \in \R^{d}$, $k$ latent attributes $\u{} \in \R^{k}$, and one ($1$) possibly unobserved categorical variable $\y{} \in\labelset$ as label from label-set $\labelset = \braces{1,2,\ldots,M}$.
For example, in citation graphs, with nodes corresponding to articles and edges to citations between articles, $\x{}$ capture observed quantities such as the bag-of-words representation of the article text; and label \y{} denotes the research area of the article (e.g., `data mining' or `machine learning').
The latent attributes \u{} correspond to properties of the articles that are not captured directly by attributes \x{} or label \y{}, but that could play a role in determining which articles are connected with a citation (as captured by adjacency matrix \A) and what research area \y{} an article is deemed to belong to.
In terms of notation, to refer to the attributes of all nodes, we write 
$\X = \braces{\x{i} \in \R^{d}, i \in \indexset{n}}$ to denote the observed node attributes, 
$\U = \braces{\u{i} \in \R^{k}, i \in \indexset{n}}$ for the latent node attributes, and 
$\Y = \braces{\y{i} \in \labelset, i \in \indexset{n}}$ for the node labels.

Having defined all elements in our setting, we now define the task we address as Problem~\ref{problem:node:classification}.
\begin{problem}[Node-Classification]
Given adjacency matrix $\A$, node features $\X$, and labels $\Y_{L}$ for a subset $L\subseteq\vertexset$ of nodes, predict labels $\Y_{\vertexset-L}$ for the remaining nodes ${\vertexset-L}$ in the graph.
\label{problem:node:classification}
\end{problem}

%% file: our_approach.tex
\section{Our Approach} \label{sec:our:approach}

Our approach for Problem~\ref{problem:node:classification} is based on a probabilistic generative model (described in Section~\ref{sec:model}) and its analysis (Section~\ref{sec:analysis}). 

\subsection{Model.} 
\label{sec:model}

\begin{figure}[t!]
	\centering
	\includegraphics[width=0.35\textwidth, height=0.25\textwidth]{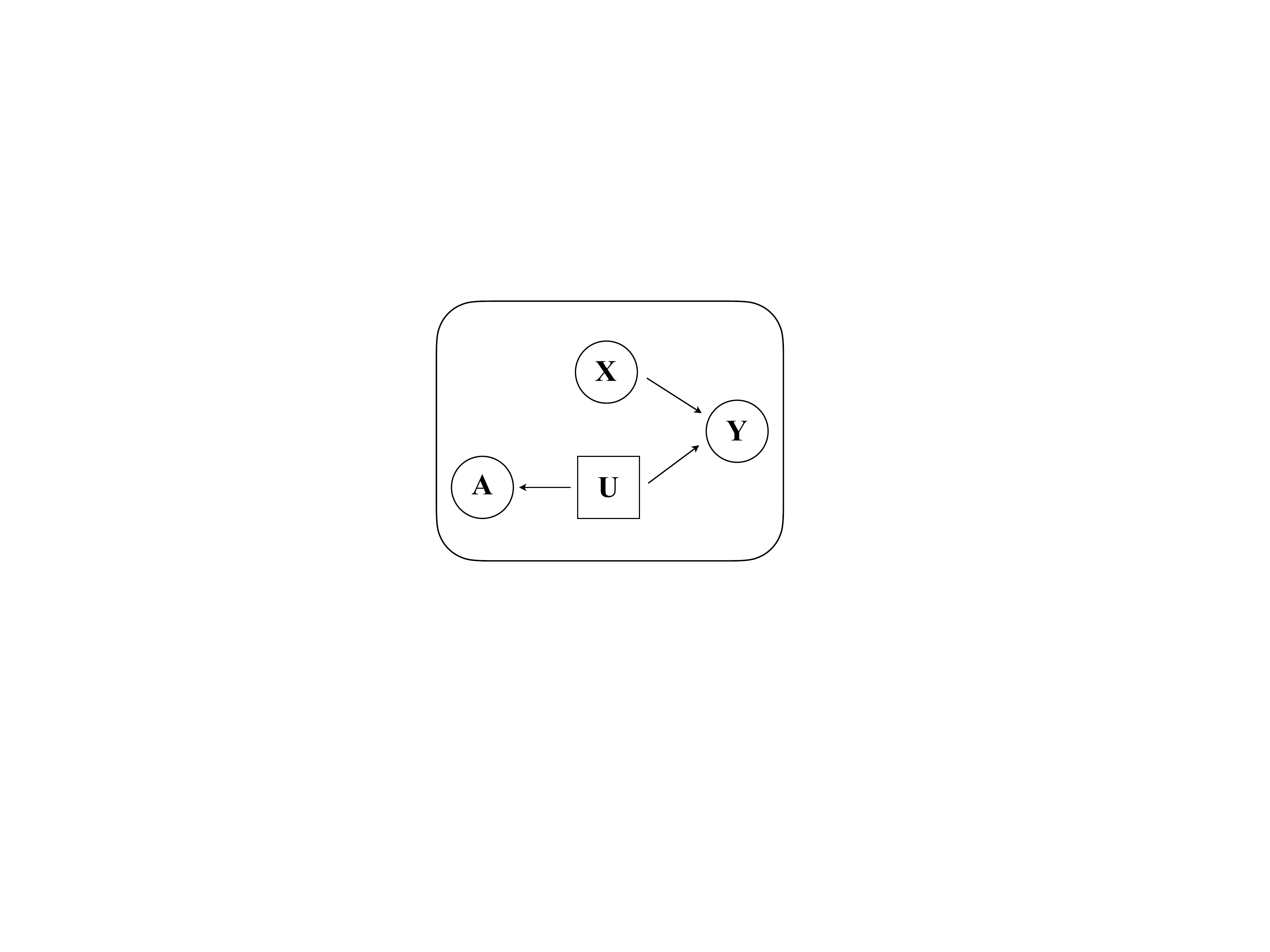}
	\caption{Visual illustration of the generative framework. Observed node attributes $\X$ (represented by circular box) and latent attributes $\U$ (represented by square box) jointly generate node labels $\Y$. The (observed) adjacency matrix $\A$ is generated from $\U$ and indirectly correlates with $\Y$ (via $\U$).}
	\label{fig:fig:X_U_A_Y}
  \end{figure}
  
Figure 2 pictorially illustrates our generative model. First, the adjacency matrix \A of the graph is generated from the latent attributes $\U$. Specifically, the probability that there exists an edge between two nodes $i$ and $j$ is given by the inverse exponent of the squares $l2$-distance between their latent attributes \u{} scaled by a factor $\scale^2$ that is set as a hyperparameter.

\begin{equation} \label{eq:e_from_u}
  \text{Pr}\squares{~ \inedgeset ~ | ~ \u{i}, \u{j} ~} = p_{ij} = \expuiujnbsigma
\end{equation}

This equips our model with the desirable property, common in many types of graph embeddings, that the closer the two nodes are in the Euclidean space of $\U$, the higher the likelihood that they are connected in the graph -- and vice versa. Therefore, \U represents a low-dimensional Euclidean embedding of the graph that preserves connectivity in the form of Eq.~\ref{eq:e_from_u}. Moreover, since the existence of an edge is independent across pairs of nodes in this model, we have

\begin{equation} \label{eq:A_from_U}
  \text{Pr}\squares{\A | \U} = \prod_{\inedgeset} p_{ij} \times
  \prod_{\notinedgeset} \round{1 - p_{ij}}.
\end{equation}

\hide{
There exist several other variations for generating graphs such as $\epsilon$-neighbourhoods wherein nodes $i, j$ are connected by an edge if $\normuiuj \leq \epsilon$ \cite{belkin2003laplacian}. However, this method can lead to several connected components. Moreover, Equation~\ref{eq:A_from_U} is in line with typical assumptions in spectral graph theory \cite{chung1997spectral}.
}

Second, node labels \Y are generated from $\X$ and $\U$. This assumption provides two benefits: (i) it allows labels to be determined by node attributes \X (directly) as well as graph structure \A (indirectly, via \U); and (ii) it allows us to express and train directly the function of the conditional distribution $\text{Pr}\squares{\Y|\X,\U}$, which we then employ for {\it node classification}, i.e., to predict unobserved node labels.

In this work, we assume that this conditional probability is given by a simple two-layer neural network,
\begin{equation} \label{eq:forward_model}
  \text{Pr}\squares{\Y|\X,\U,\WW} = \softmax\parenthesis{\text{ReLU} \parenthesis{\squares{\X\U}\W{0}} \W{1} } 
\end{equation}
where \softmax denotes the softmax function and weight matrices $\WW = \{\W{0}, \W{1}\}$ are parameters that control the effect of $\X$ and $\U$ on labels \Y. 

The reason for this choice is that we found this model to be sufficiently expressive for our empirical evaluation -- and its analysis could easily extend to more complex models (e.g., neural networks with more hidden layers).

\hide{
Finally, the probabilities above are combined into the following formula.
\begin{equation}\label{eq:gen_rule}
  \text{Pr}\squares{\Y,\A ~|~ \U; \X} = \text{Pr}\squares{\Y ~|~ \U;\X} \times \text{Pr}\squares{\A ~|~ \U}
\end{equation}
}

\subsection{Algorithms} 
\label{sec:analysis}

Problem~\ref{problem:node:classification} asks for predictions for $\Y_{\vertexset-L}$ given the data $\data = (\X, \A, \Y_L)$ that are provided as input.
Towards this end, we treat the remaining quantities in the model, i.e., the latent variables \U and the weights of the neural network \WW, as unobserved parameters \params = (\U, \WW) -- and use their maximum lilekihood values \Uhat and \What in making the predictions.
To summarize, our approach \algo proceeds in two steps: first, a training step from which we learn the maximum likelihood estimates \Uhat and \What; second, a prediction step, in which we use the learned estimates to predict the missing labels.

\subsubsection{Training} \label{sec:training}

From the product rule of probability, we have that the likelihood of the data \data for given parameters \params is proportional to

\begin{align}
    \likelihood & = \prob{\data | \params} \\
				& = \prob{\X, \A, \Y_L | \U, \WW}
				\propto \prob{\Y_L | \X ; \U, \WW} \prob{\A | \U} \nonumber
\end{align}

Algorithm~\ref{algo:train:nofle} describes the procedure to identify parameters \params that lead to maximum log-likelihood.
\begin{align}
  \paramshat & = \argmax_{\params}\log\likelihood \nonumber \\
   & = \argmax_{\U,\WW} \big(\log\prob{\Y_L | \X ; \U, \WW} + \log\prob{\A | \U}\big) \label{eq:log:likelihood}
\end{align}
In summary, Algorithm~\ref{algo:train:nofle} uses gradient descent on negative log-likelihood ($-\log\likelihood$) to alternatingly update \Uhat and \What towards their maximum-likelihood estimates, starting from an initial estimate of \Uhat given by the spectral embedding of \graph.

We provide the details below.

\spara{Choosing an initial estimate of $\U$.} 
Algorithm~\ref{algo:train:nofle} begins its iterative updates from \Uhat.
To obtain an initial maximum-likelihood estimate for it, it considers only the second term of log-likelihood (Eq.~\ref{eq:log:likelihood}), as it involves only \U and not \WW.
That is, the initial estimate \Uhat is unsupervised and (temporarily) ignores labels.
\begin{equation} \label{eq:nll_mle}
  \text{initial}\ \Uhat = \argmin_{\text{U}} \squares{-\log{\text{Pr}\round{~\A ~| ~\text{U}~}}}
\end{equation}
Using Eq.~\ref{eq:A_from_U}, we get
\begin{multline}\label{eq:general_MLE_solve}
  \begin{split}
    \Uhat & = \argmin_{\text{U}} \sum_{\inedgeset} \normuiujnbsigma - \log{\round{\prod_{\notinedgeset} \round{1 - \expuiujnbsigma}}} \\
      & \leq \argmin_{\text{U}} \sum_{\inedgeset} \normuiujnbsigma + \sum_{\notinedgeset} \expuiujnbsigma.
  \end{split}
\end{multline}
The last inequality holds because $-\round{\log{\round{1-x}}} \leq x, x \in [0, 1)$. For an appropriate choice of scale, i.e. for smaller values of $\scale$, the second term in Equation~\ref{eq:general_MLE_solve} tends to 0 and the value of the first term dominates expression~\ref{eq:general_MLE_solve}.
Given that $\laplacian$ is symmetric and $d_i = \sum_{j} \adj{i}{j}, \forall i \in \squares{n}$, based on the results of Belkin, et al \cite{belkin2003laplacian},
\begin{equation*} \label{eq:min_solve}
  \begin{split}
    \min_{\substack{\text{U}: \forall l \in [k], \, \\ \norm{\unb{l}} = 1, \\ \sum_p \unb{l_p} = 0}} \quad  \sum_{\inedgeset} \normuiujnb
      & = \text{tr}\round{\text{U}^{\text{T}} \laplacian \text{U}} = \sum\limits_{l = 1}^{k} \eigval{l}{\laplacian}.
\end{split}
\end{equation*}
The condition $\text{U}: \forall l \in [k], ~~ \norm{\unb{l}} = 1, \, \sum_p \unb{l_p} = 0$ normalizes the columns of $\text{U}$, removes translational invariance, and centers the solution around $\mathbb{0}$. This result implies that the minimum value of the first term in Equation~\ref{eq:general_MLE_solve} is the sum of the $k$ smallest eigenvalues of the graph Laplacian. 
Moreover, this minimum value is achieved when columns of $\text{U}$ are the corresponding eigenvectors. Therefore, we set as initial estimate:
\begin{equation} \label{eq:argmin_solve}
\Uhat = \squares{\eigvecExp{1}{\laplacian}, \eigvecExp{2}{\laplacian}, \ldots, \eigvecExp{k}{\laplacian}}.
\end{equation}
where \eigvecExp{i}{\laplacian} is the $i$-th smallest spectral eigenvector.

\input{algorithm}

\spara{Maximum Likelihood Estimation.}
During each training epoch, \algo first concatenates the current estimate $\Uhat$ with $\X$ and uses it as input to the neural network (Eq.~\ref{eq:forward_model}).
Since \What is present only in the first term of Equation~\ref{eq:log:likelihood}, the loss function is
\begin{equation} \label{eq:cross_entropy_error}
  \loss = \log\prob{\Y_L | \X ; \U, \WW}
\end{equation}
and the estimate \What is updated using stochastic gradient descent and standard backpropagation over it.

Subsequently, Algorithm~\ref{algo:train:nofle} treats \What as fixed and updates its estimate $\Uhat$ using the gradient of both terms of Equation~\ref{eq:log:likelihood}. 
The gradient of the first term \loss ~w.r.t. the current $\Uhat_l$ ($l$-th latent attribute) is obtained using backpropagation as follows,
\begin{multline} \label{eq:grad_loss}
  \doh{\loss}{\Uhat_{l}} = \sum\limits_{i \in \Y_L} \round{\sum\limits_{r=1}^{M} \round{\a{ir}{1} - \Y_{ir}} \W{1}_{rl}} \times \\ \round{\frac{\a{il}{0}}{1 + \exp{\round{-\z{il}{0}}}}} \times \round{\sum\limits_{p=1}^{d+k} \squares{\W{0}_{li}}}
\end{multline}
where $\a{}{0}$, $\a{}{1}$ are activations from the hidden and output layers, and $\z{}{0}$ is the weighted sum from the input layer. The index $i$ ranges over all nodes with known labels, and $r$ indexes over the $M$ different classes available for prediction.
The gradient of the second term is given by the following equation.
\begin{multline}\label{eq:grad_adj}
  \doh{\round{- ~ \log{\text{P}\round{~\A ~| ~\Uhat~}}}}{\uhat{i}} = \round{\sum_{j: \inedgeset} 2 \times \frac{\uhat{i} - \uhat{j}}{\scale^2}} + \\ \round{\sum_{j: \notinedgeset} -2 \times \frac{\uhat{i} - \uhat{j}}{\scale^2} \times \text{e}^{-\frac{\norm{\uhat{i} - \uhat{j}}^2}{\scale^2}}}
\end{multline}
where indices $i$ and $j$ range over all nodes in $\vertexset$.
Algorithm~\ref{algo:train:nofle} uses the two gradients above with corresponding learning rates $\eta_1, \eta_2$ to update \Uhat.

\subsubsection{Prediction} \label{sec:prediction}

Given maximum-likelihood estimates \Uhat and \What, we predict labels $\hat{\Y}_{\vertexset-L}$ for all nodes in $\vertexset-L$ using the softmax function applied row-wise.
\begin{equation}
  \hat{\Y}_{\vertexset-L} = \argmax_{r \in M} \softmax\round{\text{ReLU} \round{\squares{\X\Uhat}\Wh{0}} \Wh{1} }_r
\end{equation}

%% file: algorithm.tex
\renewcommand{\algorithmcfname}{Algorithm}

\begin{algorithm}[tb]

\nl {\bf Input}: \A; \X; $\Y_{L}$\\
\nl {\bf Output}: \Uhat \What (max-likelihood-estimates)  \\
\nl {\bf Parameters}: $k$ (dimensionality of $\U$); $T$ (training epochs); $\eta_1, \eta_2$ (learning rates)\\

\tcc*[h]{Initialize $\Uhat$ with first spectral eigenvectors}\\
\nl $\laplacian \gets \diagmat - \A$ \\
\nl $\Uhat \gets \squares{\eigvecExp{1}{\laplacian}, \eigvecExp{2}{\laplacian}, \ldots, \eigvecExp{k}{\laplacian}}$ \\

\For{$t \gets 1 ~ \text{\textbf{to}} ~T$}{
	\tcc*[h]{Update \What}\\
	\nl $\What \gets \text{backpropagation}(\squares{\X ~ \Uhat},\What,\Y_{L})$
	\tcc*[h]{Update \Uhat}\\
	\nl $\Uhat \gets \Uhat - \eta_1 \cdot \doh{\loss}{\Uhat} - \eta_2 \cdot \doh{\round{- \log{\text{Pr}\squares{~ \A ~ | ~ \U ~}}}}{\Uhat}$
}
\caption{\algo}
\label{algo:train:nofle}
\end{algorithm}

%% file: experiments_synthetic_data.tex
\section{Experiments} \label{sec:experiments}

In this section, we empirically evaluate the performance of \algo on synthetic and real-world datasets.

\spara{Baselines.} We evaluate \algo against standard baselines from a variety of different design philosophies.
\squishlist
  \item \textit{\algo and variants:} \algo-NU wherein we do not update the initial estimate of $\Uhat$ w.r.t. labels or graph structure during training, and \algo-R where the initial estimate of U is a random matrix. 
  \item \textit{Graph-structure based approaches:} Label Propagation (LP) \cite{zhu2002learning}, and DeepWalk (DW) \cite{Perozzi2014} that encodes neighbourhood information via truncated random walks. These do not incorporate node attributes.
  \item \textit{Deep attributed embeddings:} We evaluate LANE \cite{huang2017label} which constructs node embeddings that encode graph structure and node attribute information, in addition to node labels.
  \item \textit{Graph-convolution based approaches:} GCN \cite{Kipf2016} and GraphSAGE (mean aggregator) as representatives of graph convolutional networks. We acknowledge that this is extremely active area of research today and there are several approaches that demonstrate improved performance along axes such as training efficiency~\cite{cong2020minimal}, explainability~\cite{wu2019net}, etc. on various real-world datasets \cite{hu2020open}. For economy of space, we empirically compare with two benchmark, representative methods to demonstrate our central point: these approaches sometimes fail because they strictly and inherently enforce homophily and social influence.
\squishend

We were unable to reproduce the node classification results for AANE \cite{huang2017accelerated} based on the available implementation\footnote{\url{https://github.com/xhuang31/AANE_Python}}. Further, we could not locate the authors' implementations for DANE \cite{gao2018deep}. Therefore, we do not report these results.\\

\spara{Experimental Setup.} We implement \algo, \algo-NU, and LP in Pytorch. We use out-of-box Pytorch implementations of DeepWalk\footnote{\url{https://github.com/phanein/deepwalk}}, GCN\footnote{\url{https://github.com/tkipf/pygcn}} and GraphSAGE\footnote{\url{https://github.com/williamleif/graphsage-simple}}. And, we use a MATLAB implementation\footnote{\url{https://github.com/xhuang31/LANE}} of LANE \cite{gao2018deep}. In all of our experiments, as is standard, all approaches receive only the adjacency matrix of the graph $\A$ and the node attributes $\X$ as input, along with the same 10\% and 20\% of the node labels for training and validation, respectively. Wherever available, we use the default hyperparamater configurations as suggested in the original papers. Otherwise, we grid search over the hyperparameter space to find the best setting for all of our baselines. We perform all experiments on a Linux machine with 4 cores and 32GB RAM.

\spara{Reproducibility.} To aid further research, we make our code publicly available\footnote{\url{https://version.helsinki.fi/ads/jane}}. 
This also includes an implementation for constructing synthetic datasets as described below. The real-world datasets (cf. Section~\ref{subsec:experiments_real_data}) are publicly available.

\subsection{Node Classification on Synthetic Data} \label{subsec:experiments_synthetic_data}

The goal of these experiments is two-fold -- (1) demonstrate fundamental differences between and limitations of existing classification approaches using synthetic datasets wherein labels derive from only $\X$, only $\U$, or partly from both, and (2) show the strengths and general-purpose nature of \algo vis-à-vis the source of node labels.

\begin{figure*}[ht!]
	\captionsetup[subfigure]{font=scriptsize,labelfont=scriptsize}
	\centering
	\begin{subfigure}[t]{\textwidth}
	  \includegraphics[width=\textwidth]{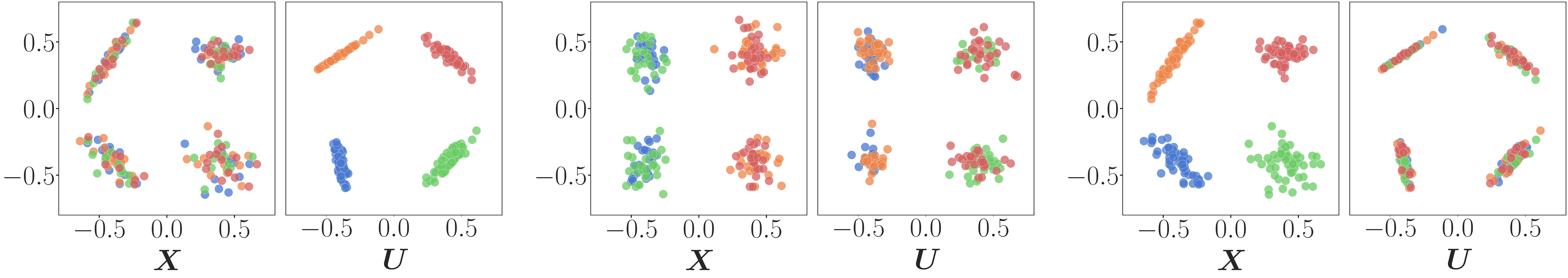}
		\caption{Observed and latent attributes $\X$ and $\U$ (respectively) generated for $\alpha=0.0$ (left), $\alpha=0.5$ (center), and $\alpha=1.0$ (right).}
		\label{subfig:XU}
	\end{subfigure}
	\begin{subfigure}[t]{\textwidth}
	  \includegraphics[width=\textwidth]{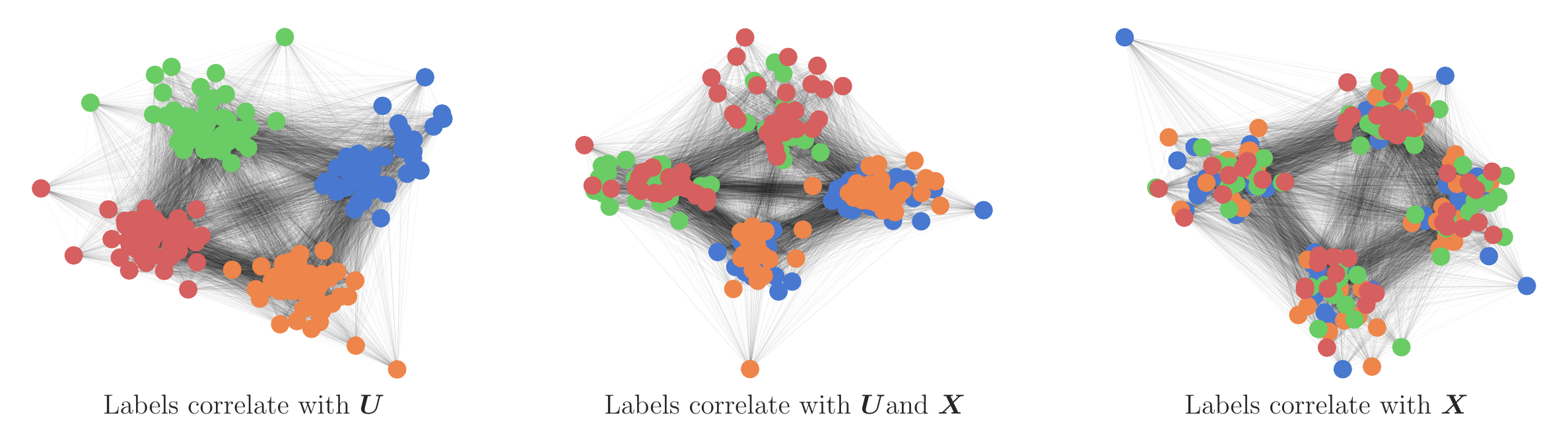}
		\caption{Random graph generated from latent attributes $\U$ for $\alpha=0.0$ (left), $\alpha=0.5$ (center), and $\alpha=1.0$ (right).}
		\label{subfig:G}
	\end{subfigure}
	\begin{subfigure}[t]{\textwidth}
	  \includegraphics[width=\textwidth]{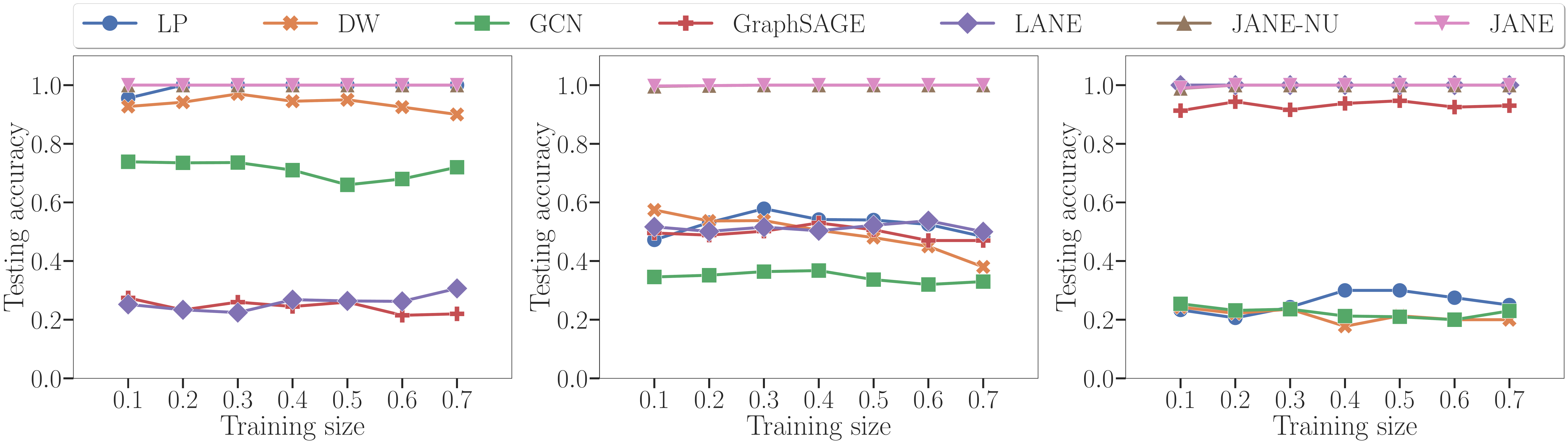}
		\caption{Node classification accuracy averaged over randomly generated datasets for $\alpha=0.0$ (left), $\alpha=0.5$ (center), and $\alpha=1.0$ (right).}
		\label{subfig:acc_train_size_alpha}
	\end{subfigure}
	\caption{Figure \ref{subfig:XU} depicts 3 synthetically generated datasets $\X$ and $\U$ with class labels \Y influenced only by $\U$ ($\alpha = 0.0$), partly by $\U$ and partly by $\X$ ($\alpha = 0.5$), and only by $\X$ ($\alpha = 1.0$). Figure \ref{subfig:G} shows an instance of the corresponding graph generated from $\U$ according to Equation~\ref{eq:A_from_U}. Figure \ref{subfig:acc_train_size_alpha} compares the node classification accuracy of \algo and \algo-NU with the baselines averaged over 5 random train-test splits.
	}
	\label{fig:experiments_synthetic_data}
  \end{figure*}

\spara{Synthetic Datasets.} Figure~\ref{fig:experiments_synthetic_data} describes representative synthetic datasets generated according to the framework described previously.

We set the number of individual node features $|\X| = d = 2$ and number of latent features $|\U| = k = 2$. We generate these features (gaussian-distributed) for $n=200$ points, each of which belongs to one of $M=4$ classes and set the scale $\scale^2 = 1$.
An influence parameter, $\alpha\in\squares{0, 1}$, controls the degree to which node labels derive from \X or \U: $\alpha = 0.0$ signifies that they derive only from $\U$ and are independent of $\X$; $\alpha = 1.0$ that they derive only from $\X$ and are independent of $\U$; and $\alpha = 0.5$ that they derive equally from \X and \U (specifically, without loss of generality, only the first feature from $\X$ and $\U$ contributes to label assignment).
Figure~\ref{subfig:XU} depict an instance of $\X$ and $\U$ each for $\alpha = \braces{0.0, 0.5, 1.0}$, respectively. The colors of points represent classes. We construct the adjacency matrix from $\U$ as per Equation~\ref{eq:A_from_U}. Figure~\ref{subfig:G} depicts an instance of the corresponding graphs. 

\spara{Implementation Details.} We use Scikit-Learn's ~\cite{scikit-learn} \textsc{make\_classification} to generate these datasets. Approaches do not have access to $\alpha$ or $\U$. \algo is trained as a two-layer neural network for a maximum of $T=200$ epochs with dropout of 0.2 for each layer, weight decay of $5e^{-2}$, and learning rate of 0.005 using Adam. We set the number of eigenvectors $k=2$ and choose a scaling factor $\scale^2=0.01$.

\spara{Performance.} Figure~\ref{subfig:acc_train_size_alpha} shows performance as a function of increasing training set sizes.

\squishlist
  \item $\alpha = 0.0$: LP and DW infer that labels derive from $\A$ (indirectly). GCN converges attribute values of nodes in the same cluster but is not perfectly accurate because $\X$ does not correlate with $\Y$. LANE forces the proximity representation to be similar to the attribute representation and then smoothens it using the labels. It does not perform well since there is no correlation between them. 
  \item $\alpha = 0.5$: LP, DW are able to correctly classify nodes belong to 2 out of 4 classes, i.e. precisely those nodes whose labels are influenced by $\U$. Conversely, LANE is able to classify those nodes belong to two classes of nodes that correlate with $\X$. GCN smoothens attribute values of adjacent nodes and thus can correctly infer labels correlated with $\X$.
  \item $\alpha = 1.0$ LP and DW reduce to random classifiers since adjacent nodes do not have similar labels. GCN reduces to a nearly random classifier because by forcing adjcent nodes with different attribute values to become similar, it destroys the correlation between $\X$ and the labels. 
\squishend

In each of the three cases, \algo-NU and \algo achieve perfect accuracy because they flexibly learn during training whether labels correlate with $\X$, $\A$ (indirectly), or partially both. While these datasets are simplistic in nature, this demonstrates how the homophily assumption--requiring nodes with similar proximity and attributes to have the same labels--limits the performance of other approaches.

%% file: experiments_real_data.tex
\subsection{Node Classification on Real-World Data} \label{subsec:experiments_real_data}

We seek to understand -- (1) to what extent \algo can capture more complex network structures and their correlations with node labels, and (2) how well \algo compares with our baselines on these datasets.

\spara{Datasets.} We show our results on four tasks for a total of 9 graphs. If the original graph is diconnected, we extract node attributes and labels belonging to its largest connected component. 
\squishlist
	\item \textit{Citation Networks:} We use Cora, Citeseer, Pubmed~\cite{sen2008collective}, and UAI2010 \cite{wang2018unified}. Here, nodes represent academic papers, edges denote a citation between two nodes, node features are 0/1-valued sparse bag-of-words vectors and class labels denote the subfield of research that the papers belong to.
	\item \textit{Social Networks:} We focus on BlogCatalog and Flickr where the task is to predict pre-defined categories of blogs and images, respectively. Nodes are users that post content, edges represent follower relationships, and features are specified by a list of tags reflecting the interests of the users \cite{li2015unsupervised}.
	\item \textit{Air-traffic Networks:} Based on flight records from Brazil, Europe, and USA, each node is an airport and an edge indicates a commercial airline route exists between them. Labels denote the level of activity in terms of people and flights passing through an airport \cite{ribeiro2017struc2vec}. Since no features for the nodes exists, we assign the all-ones vector as the sole attribute.
	\item \textit{Biological Networks:} We use a processed protein-protein interaction (PPI) dataset \cite{Hamilton2017a} where the task is to identify protein roles based on gene ontology sets using positional gene sets, motif gene sets, and immunological signatures as features \cite{subramanian2005gene}.
\squishend
Table~\ref{table:dataset_statistics} summarizes the dataset statistics.

\begin{table}[tb!]
	\setlength\tabcolsep{2pt}
	\fontsize{9}{10}\selectfont
	\centering
    \begin{tabular}{lcccc}
      \toprule[1.5pt]
	      {\bf Dataset} &
				\begin{tabular}{c@{}} \textbf{Nodes} \end{tabular} &
				\begin{tabular}{c@{}} \textbf{Edges} \end{tabular} &
				\begin{tabular}{c@{}} \textbf{Classes} \end{tabular} &
				\begin{tabular}{c@{}} \textbf{Features} \end{tabular} \\
      \midrule
				Cora	&	2708	&	5429	&	7	&	1433 \\
				Citeseer &	3327	& 4732	& 6	& 	3703 \\
				Pubmed & 19717 & 44325 & 3 & 500 \\
				UAI2020 & 3067 & 56622 & 18 & 4973 \\
				BlogCatalog & 5196 & 171743 & 6 & 8189 \\
				Flickr & 7575 & 239738 & 9 & 12047 \\
				Brazil & 131 & 1038 & 4 & NA \\
				Europe & 399 & 5995 & 4 & NA \\
				USA & 1190 & 13599 & 4 & NA \\
				PPI & 2373 & 56952 & 1 & 121 \\
      \bottomrule[1.5pt]
    \end{tabular}
	\caption{Summary of dataset statistics.}
	\label{table:dataset_statistics}
\end{table}

\spara{Experimental Setup.} for the citation datasets, we use the same train-validation-test splits as in Yang, et al \cite{Yang2016} minus the nodes which do not belong to the largest connected component. These comprise of 20 samples for each class and represent 5\% of the entire dataset. We use 500 additional samples as a validation set for hyperparameter optimization as per Kipf, et al \cite{Kipf2016} to enable fair comparison. For all other tasks, we use 10\% and 20\% of the dataset for training and validation, respectively. We evaluate the performance of all approaches on the remaining nodes of the graph. Values of hyperparameters $k$, the number of eigenvectors of the graph Laplacian, and the scaling factor $\scale$ are determined empirically so as to minimize Equation~\ref{eq:general_MLE_solve}.

\begin{table*}[thb!]
	\setlength\tabcolsep{2pt}
	\fontsize{7}{8}\selectfont
	\centering
	\begin{tabular}{lccccc||ccc}
      \toprule[1.5pt]
	      {\bf Dataset} &
				\begin{tabular}{c@{}} \textbf{LP} \end{tabular} &
				\begin{tabular}{c@{}} \textbf{DW} \end{tabular} &
				\begin{tabular}{c@{}} \textbf{GCN} \end{tabular} &
				\begin{tabular}{c@{}} \textbf{GraphSAGE} \end{tabular} &
				\begin{tabular}{c@{}} \textbf{LANE} \end{tabular} &
				\begin{tabular}{c@{}} \textbf{\algo-R} \end{tabular} &
				\begin{tabular}{c@{}} \textbf{\algo-NU} \end{tabular} &
				\begin{tabular}{c@{}} \textbf{\algo} \end{tabular} \\
      \midrule
				Cora & 74.29 & 32.54 & 79.20 & \textbf{79.70} & 64.50 & 59.34 & 78.87 & \textbf{79.24} \\
				Citeseer & 67.19 & 61.31 & 69.11 & \textbf{69.84}  & 56.93 & 52.80 & 69.71 & \textbf{69.77} \\
				Pubmed & 64.33 & 77.91 & \textbf{82.26} & 81.86 & 78.81 & 72.91 & 82.70 1 & \textbf{83.18}\\
				UAI2010 & 42.18 & 44.67 & 49.52 & 61.22 & 66.36 & 45.96 & 66.91 & \textbf{69.84}\\
				BlogCatalog & 48.81 & 37.18 & 70.53 & 75.13 & \textbf{82.26} & 70.85& 74.85 & 77.53 \\
				Flickr & 41.24 & 33.61 & 49.58 & 57.17 & \textbf{64.12} & 49.24 & 59.92 & 61.68 \\
				Brazil & 26.83 & 51.71 & 27.21 & 28.26  & 24.59 5 & 29.32 & 68.55 & \textbf{69.93} \\
				Europe & 26.33 & 47.15 & 25.46 & 46.61  & 27.50 & 28.81 & 45.97& \textbf{50.21} \\
				USA & 24.73 & 56.67 & 24.13 & 48.63  & 25.74 5 & 27.67 & 57.10 & \textbf{61.91} \\
				PPI & 38.83 & 47.19 & 51.19 & \textbf{61.26} & 53.73 & 45.12 & 62.68 & \textbf{62.98}\\
      \bottomrule[1.5pt]
    \end{tabular}
	\caption{Classification accuracy (\%) on test data averaged over 10 independent runs of $T=200$ epochs each. Bold denotes best average accuracy or overlapping with best accuracy range (within $\pm 0.3$ standard deviation). \algo-NU and \algo perform as well as or better than LANE, GCN and GraphSAGE on most datasets and consistently well across datasets.
	}
	\label{table:realdata_accuracy_results}
\end{table*}

\spara{Performance Analysis.} Table~\ref{table:realdata_accuracy_results} provides the average accuracy of each approach over 10 independent runs for $T=200$ epochs each for GCN, GraphSAGE, LANE, and \algo.
Values in bold denote that the approach either performs best or its accuracy range overlaps with that of the best. 
\algo-NU and \algo consistently outperform LP and DW on all datasets by significant margins. 
\algo-R on the other hand, has significantly lower performance either because it gets stuck in a local optima or it requires much longer training time. This shows that the choice of the initial estimate is crucial to performance. The embedding obtained from DW is unsupervised and thus its predictive power is limited in comparison to our $\Uhat$ which is label-informed. Both variants of \algo are competitive with GCN and GraphSAGE on Cora, Citeseer, and UAI2020. For instance, GraphSAGE achieves 79.70\% accuracy on Cora, while \algo gets 79.24\% which is within the margin of error. 
We observe strong performance by \algo-NU and \algo on Pubmed and PPI. This may be explained in part by Li, et al \cite{Li2018}'s observation that Pubmed exhibits strong manifold structure as well as \algo's ability to better utilize graph structural information. Lastly, we find significant gains on the social, and air-traffic datasets. LANE \cite{huang2017label} report strong performance on Blogcatalog and Flickr, but its performance is noticeably poorer on citation, flights, and biological datasets. \\


\begin{figure}[h!]
  \centering
    \includegraphics[width=0.70\textwidth]{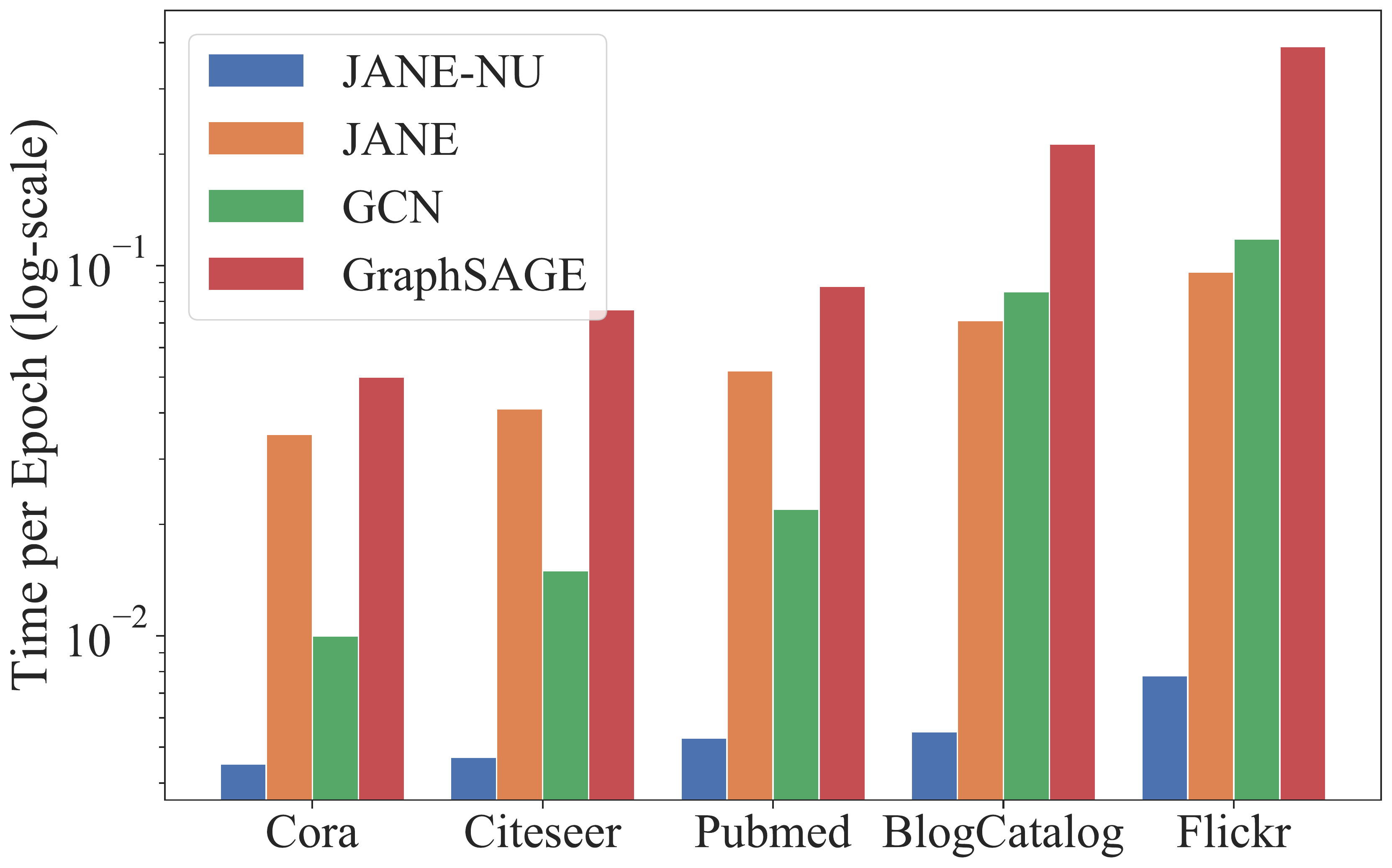}
	\caption{Average training time per epoch (seconds) of baselines on different datasets.}
    \label{fig:dataset_vs_runtime}
\end{figure}

\spara{Runtime.} Figure~\ref{fig:dataset_vs_runtime} plots the average time for a single training epoch for \algo-NU, \algo, GCN, and GraphSAGE on various datasets. \algo is comparatively the slowest model while \algo-NU is by far the fastest. This is because \algo-NU is a vanilla neural network model that does not update its estimate of $\Uhat$. Further, a single laplacian eigenvector can be approximately computed using the Lanczos algorithm in $\tilde{\mathcal{O}}\round{|\edgeset|}$ (up to log factors). Thus $k$ eigenvectors can be computed in $\tilde{\mathcal{O}}\round{k \cdot |\edgeset|}$ and this is a one-time operation. Note, we do not need to perform a full eigendecomposition. 

\begin{figure}[h!]
  \centering
    \includegraphics[width=0.50\textwidth]{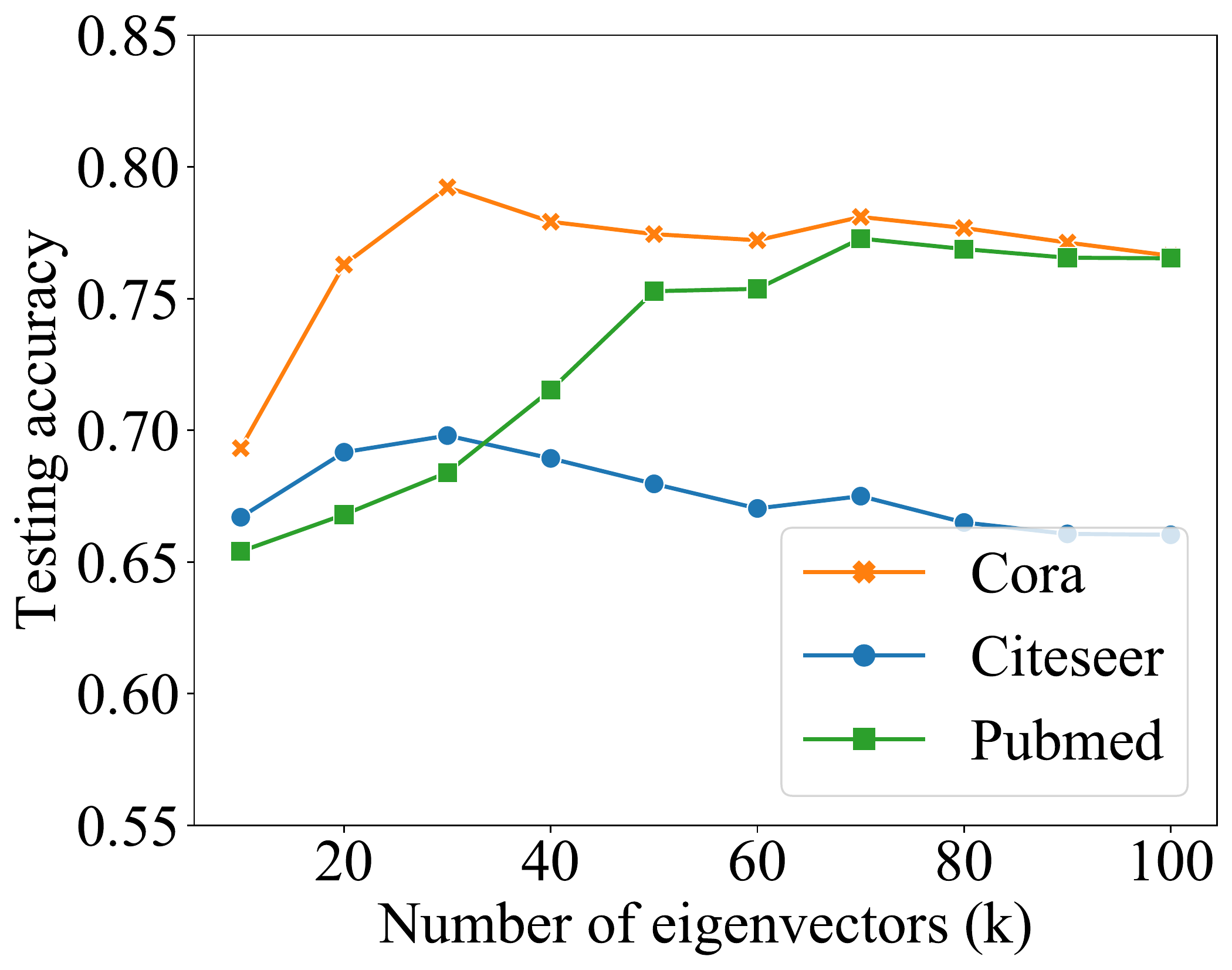}
	\caption{\algo's classification accuracy (\%) on test data averaged over 10 runs w.r.t. number of eigenvectors.}
    \label{fig:n_eigs_vs_accuracy}
\end{figure}

\spara{Parameter Sensitivity.} A crucial parameter of \algo is the number of eigenvectors used as features during training. Figure~\ref{fig:n_eigs_vs_accuracy} demonstrates their impact on performance. In each case, \algo is allowed a maximum of 200 training epochs. We find that test accuracy consistently increases as dimension of $\Uhat$ increases up until a certain threshold. However, increasing beyond this threshold introduces noise and reduces performances. Having too many eigenvectors gives marginal and diminishing returns in performance while increasing the runtime. 

\spara{Limitations.} The primary limitation of \algo is the time and memory requirement for computing the gradient of $\A$ w.r.t. $\Uhat$ (cf. Equation~\ref{eq:grad_adj}). These requirements grow linearly in graph size. Since the gradient is computed in every training epoch, it may not be viable to fit it into GPU memory. Future work can outline procedures for generating mini-batches and schedules for updating the estimate of $\Uhat$ using $\A$ at regular intervals (as opposed to every epoch as we do now) whilst still gaining the same benefits. It is crucial to note however, that \algo-NU is orders of magnitude faster than the other methods while also demonstrating strong performance.

%% file: conclusion.tex
\section{Conclusion} \label{sec:conclusion}
In this paper, we developed an approach to node classification that flexibly adapts to settings ranging between graphs where labels are predicted by node proximity, on one hand, and graphs where labels are predicted by node attributes, on the other. We propose a generative framework to demonstrate how graph structural information and node attributes both, can influence the labels of nodes. Even simple instances of such situations, as shown in Figure~\ref{fig:toy_graph_illustration} and empirically evaluated in Figure~\ref{subfig:G}, severely affect the performance of various embedding methods and standard graph convolutional networks. Our principled approach, \algo, jointly and effectively utilizes Laplacian eigenmaps and individual features for classification leading to strong performance on a variety of datasets. Given its simplicity, interpretability and performance, \algo can serve as a useful starting point in designing models that holistically account for different sources of node labels. As a future direction, we aim to the evaluate the design and performance of advanced graph neural networks that go beyond requiring or enforcing homophily.